\documentclass{aa_nohead}

\usepackage{epsf}

\input epsf

\begin{document}

\thesaurus{06(08.14.2; 08.23.1; 02.03.3; 02.08.1; 02.14.1)}

\title{Three-Dimensional Simulations of Classical Novae}

\author{A. Kercek \inst{1} \and W. Hillebrandt \inst{1} \and
J. W. Truran \inst{2}}

\institute{Max-Planck-Institut f\"ur Astrophysik,
Karl-Schwarzschild-Strasse 1, D-85740 Garching, Germany
\and Laboratory for Astrophysics and Space Research, Enrico Fermi Institute,
University of Chicago, Chicago, IL 60637, USA}

\date{Received <date>; accepted <date>}

\maketitle

\markboth{A. Kercek et al.: Three-Dimensional Simulations of Novae}
{A. Kercek et al.: Three-Dimensional Simulations of Novae}

\begin{abstract}

We present first results of three-dimensional (3D-) calculations of 
turbulent and degenerate hydrogen-burning on top of a C+O white 
dwarf of 1.0 M$_{\odot}$. The simulations are carried out
by means of a code which solves Euler's
equation for an arbitrary equation of state together with a
nuclear reaction network and the energy input  from nuclear reactions    
on a Cartesian grid covering a fraction of the white dwarf's
surface and accreted atmosphere.
The flow patterns we obtain are very different from those
of earlier 2D simulations using the same initial conditions
and the same numerical resolution. The possibility of self-enrichment
of the accreted hydrogen-rich atmosphere  with carbon and oxygen
from the surface layers of the white dwarf during the violent phase 
of the burning is investigated, and it is demonstrated that
self-enrichment proceeds too slowly if the accreted gas has near-solar
CNO-abundances at the onset of the thermonuclear runaway. As a result,
we do not find a fast nova outburst. This conclusion remains valid
if the initial metallicity of the accreted gas is raised by
a factor of five. Therefore we conclude that fast nova
outbursts indeed require huge enrichments of C and O, as postulated
from spherically symmetric models, and that the mechanism which
leads to such enhancements must operate prior to the outburst.


\end{abstract}

\section{Introduction}

The standard model for the outburst of a classical nova
is a thermonuclear runaway (TNR) in the accreted hydro\-gen-rich 
envelope on top of a white dwarf in a close binary system 
(\cite{Starr89} 1989, 1993,
1995; \cite{Truran82} 1982, 1990).
Spherically symmetric models of the runaway,
based on realistic nuclear reaction rate networks and mixing-length
theory (MLT) of convection, have been investigated by many authors
(\cite{Starr74} 1974, 1985; \cite{Prialnik78} 1978;
\cite{MacD80} 1980), and the results
were in good agreement with observational data, such as the
total amount of energy released and the metallicity 
and the abundances of the expelled envelope material, provided
considerable enrichment of the atmosphere with CNO-elements
was assumed. Although the actual numbers depend 
on the mass of the white dwarf and the accretion rate, 
typical enhancements of up to a factor of 10 relative
to solar values were required for models of fast novae.

Despite of the success of these models, 
two questions remain to be answered:
\begin{itemize}
\item 1. Where does the enrichment come from?
\item 2. What is changed if one avoids the mixing-length theory of 
          convection?
\end{itemize}
In this paper we concentrate on the second question and will
eliminate one suggested answer to the first one in passing.
We shall mainly follow the arguments given in a recent work by us
(Kercek, Hillebrandt \& Truran 1998; henceforth referred to as KHT),
where these questions were addressed by direct two-dimensional
simulations. For convenience we first summarize here the numerical  
method as well as the main results obtained in KHT.

In KHT (as well as here) we performed numerical simulations based 
upon the hydro-code PROMETHEUS (Fryxell et al. 1989), a PPM-type
code with nuclear reactions included. Curvature effects were ignored and
the surface layers of the white dwarf as well as its envelope were 
represented by a plane-parallel sheet. The advantage of this
approach was that periodic boundary conditions could be used, thereby
avoiding common problems of numerical simulations of free convection,
namely that reflecting boundaries may act like a containment and may affect
the flow patterns in an unphysical way.  Effects coming from the
finite numerical resolution were investigated and it was found that,
as far as the general properties of the simulations were concerned,
a grid of moderate resolution. e.g. 220 $\times$ 100 (horizontal
$\times$ vertical) grid points, covering a total of 1800km $\times$
1000km of the white dwarf's surface and atmosphere, was sufficient
to obtain stable results for integral quantities such as the energy
generation rate and the rate of mixing of carbon and oxygen from the
white dwarf into the atmosphere.

In our version of PROMETHEUS, 
nuclear reactions are incorporated by solving 
together with the hydrodynamics a nuclear reaction network including 
12 nuclear species, i.e., $^1$H, $^4$He, $^{12}$C, $^{13}$C, $^{13}$N, $^{14}$N,
$^{15}$N, $^{14}$O, $^{15}$O, $^{16}$O, $^{17}$O, and $^{17}$F, linked
by reactions described in Wallace \& Woosley (1981). The reaction rates 
are taken from Thielemann (private communication). 
Following \cite{Mueller86} (1986), we solve the network 
equations and the 
energy source equation simultaneously to avoid numerical
instabilities.
 
Since the accretion process prior to the TNR needs several $10^5$ years
in order to get an initial model for the multi-dimensional
simulations, one has to calculate the accretion phase and the slow
stages of the burning by means of one-dimensional implicit hydro codes.
Here as well as in KHT we used a model computed to the onset of the
vehement phase of the TNR by \cite{Glasner97} (1997).
During this early phase, hydrogen-rich matter of solar composition 
(Z = 0.015) is accreted onto
the surface of a 1 M$_{\odot}$ C+O 
white dwarf at an accretion rate of 5.0$\cdot$10$^{-9}$M$_{\odot}$yr$^{-1}$. 
As the temperature rises at the bottom of the envelope
convection sets in. Convection is taken into account by the
MLT in the one dimensional model. The model provides the 
density, temperature, entropy, and composition profiles at the point at 
which the temperature at the bottom of the 
envelope reaches 10$^8$ K. The mass of the hydrogen shell is then
about $2 \cdot 10^{-5}$M$_{\odot}$ and the metallicity is now Z = 0.02.
This model was mapped onto a radial row of the 2D grid in KHT,
relaxed for several hundred dynamical time scales, and then mapped 
onto the full 2D grid. In the present work, the same procedure is
applied to a 3D grid. Finally, an initial perturbation is
superimposed upon this hydrostatic configuration.

The main features of the 2D flow fields which showed up in KHT 
were the appearance of small persistent coherent structures 
of very high vorticity (and velocity)  compared to the
background flow. Although they were not fully resolved 
they caused effects that were essentially independent
of the resolution. During the early phase of the thermonuclear runaway 
they dominated the flow patterns and resulted in very little
overshoot and mixing. At late times, after steady slow mixing and
with increasing nuclear energy production, they became
weak, but showed up again after hydrogen had mainly been burnt and the
energy generation rate dropped. The net effect was that KHT did 
find some self-enrichment, but on time-scales much
longer than in previous calculations. Moreover, for initially solar
composition of the accreted gas the rise time of the temperature 
was very long, of the order of 1000s, and peak temperatures never 
exceeded 2$\cdot$10$^8$K. Therefore these models did not resemble the
properties of a fast nova.

Of course, the question arises if these conclusions would remain
valid if the assumption of 2D symmetry were drop\-ped. First, one does
not expect to find the small-scale stable vortices in 3D because
they are a property of plane-parallel 2D flows, as was discussed
in KHT. On the other hand, one also expects the dominant
linear scale of convective eddies to be smaller in 3D than in 2D 
and, therefore, mixing and convective overshooting could be even
less efficient than was found by KHT. 

In the following sections we will answer this question. In Sect. 2
we briefly describe the parameters of the models we have computed.
In Sect. 3 we present and discuss the results of two sets of
computations, one with solar initial composition of the accreted gas
and a second with 5 times higher metal abundances. Some conclusions
and an outlook follow in Sect. 4. We will concentrate on the
astrophysical aspects of our simulations and leave results and
questions related to the physics of turbulent combustion in the
so-called well stirred regime to a forthcoming paper.

\section{The models}

The computational grid of our present computations covers a fraction
of 1000km (vertical) $\times$ 1800km $\times$ 1800km (lateral) of
the white dwarf's surface and atmosphere
by 100 $\times$ 220 $\times$ 220 grid points in a Cartesian mesh.
This is equivalent to the ``low resolution'' case of KHT. The lateral
grid is equidistant whereas in vertical direction we use a
non-equidistant grid with the highest spatial resolution (5km) in
the white dwarf's surface and the bottom layers of its atmosphere.
As an initial perturbation we increase the temperature by 1\% in
20 zones right on top of the white dwarf. In contrast to KHT, we
choose a somewhat larger perturbed area in order to save
computer time. This, however, is not a problem since the initial
conditions are a bit artificial anyhow, and we saw in our
previous 2D simulations that information about the initial
perturbation is lost after about 15s.

Since we fix the initial conditions according to the model of 
\cite{Glasner97} (1997) the only parameter which still can be varied
within reasonable limits is the initial metallicity of the atmosphere. 
But, because solving the reactive Euler-equations in 3D with realistic
equations of state is ``expensive'' (on 512 processors of a CRAY
T3E one time-step needs about 8s CPU-time, and about 2$\times$10$^5$
time-steps are required per simulation), we cannot perform an
extensive parameter study. We decided rather to choose solar 
composition for a first simulation (in order to compare the results
with our earlier 2D computations) and to increase the metallicity
to 10\% in a second run. This second simulation is meant to 
investigate whether or no moderately higher CNO abundances 
lead to significantly more violent nuclear burning, thereby
changing a ``slow'' nova to a fast one. Although it is certainly of
interest to increase C and O by another factor of 2 to 3 in order to 
match the conditions which would give a fast nova in 1D simulations,
we did not do it because of obvious inconsistencies of our initial 
model in that case. 

Combustion in novae is peculiar because convective overturn times are
much shorter than the nuclear reaction time-scale. Therefore, in general,
stirring proceeds much faster then nuclear burning. In our
approach to this problem, namely by direct numerical simulations,
we resolve the large scale inhomogeneities and velocity
fluctuations well, but the small scale dissipation and mixing is
dominated by numerical discretization effects. This is not a problem
as far as the energy transport by convective motions is concerned
because the energy flux is predominantly carried by the largest
eddies. Mixing by turbulent motions, however, happens on very small
scales which are definitely not resolved in our
computations. Therefore we will overestimate the mixing due to the
numerical diffusion of matter on the grid scale. But since, as we
shall show, even with our treatment mixing is too slow for a fast nova
to occur, direct simulations are a conservative approach to the
problem under consideration here. 

\section{Results of the 3D simulations}

\subsection{The case of near-solar metallicity (Z=0.02)}

In this subsection we present and discuss the results we have
obtained for an atmosphere of nearly solar metallicity, the main emphasis
being on a comparison with KHT. The initial abundances were obtained
by evolving the accreted atmosphere of a given metallicity through
steady hydrogen burning to the vehement stage of the TNR.  
We shall first analyse the evolution of the convective flow fields,
and then discuss integral quantities such as the laterally averaged
temperature and the nuclear composition. We computed this model over
about 400s.

\subsubsection{The evolution of the flow fields}

In order to visualize our results we present them in the form of
2-dimensional cuts and iso-surfaces in three dimensions. The
2D-cuts show, colour-coded, the absolute values of the velocity as a
function of time. They indicate where in the atmosphere of the white
dwarf most of the convective transport is happening and can easily be 
compared with our earlier 2D results. In contrast, the iso-surfaces of
the absolute values of the velocity field at fixed times are used to
discuss the characteristic structure and scales of the convective 
eddies.  
 
Figure \ref{flow-3D} gives the temporal evolution of the velocities for a typical 2D
cut. As in the 2D simulations of KHT, the surface layer is ignited by
an ``ignition string'' perpendicular to the plane shown. After a
couple of seconds the entire surface layer is burning and the first
convective eddies form. This stage is reached a little earlier than in
the 2D simulations, mainly because a larger volume was perturbed
initially. From then on the evolution is very different from the 2D
case. No regular structures appear. In contrast, the flow field is very
irregular. Violent eruptions occur occasionally at later times, followed
by periods of rather quiet burning. Towards the end of the computations,  
convective motions extend all the way into the upper atmosphere.

The evolution of the convective patterns can be seen more clearly in
the iso-surface plots Figs. \ref{vel-ata-3D} to \ref{vel-cza-3D}. 
Figure \ref{vel-ata-3D}, taken at 50s, shows that
most of the fast eddies in the beginning are confined to a narrow
layer near to the white dwarf's surface, extending over roughly 150km.
This burning layer is well separated from an upper convective zone
with large but slow eddies. But even the convective motions near the
surface are slow in comparison to the sound velocity which is above
10$^8$cm/s there. During this phase, the burning matter does not expand
significantly and also the upper atmosphere is only moderately heated
from below. Consequently, the temperature rises in the burning region
above 10$^8$K, and the burning becomes more violent. As is shown
in Fig. \ref{vel-bmm-3D}, after about 100s, convective velocities begin to exceed
10$^7$cm/s and some eddies start to penetrate into the upper atmosphere. From then on
very efficient heat transport by convection leads to an expansion
of the atmosphere, and energy going into lift-off, and adiabatic energy losses 
balance essentially
gains from nuclear burning. After 200s, finally, the entire atmosphere
is well stirred by convective eddies of all sizes with typical
velocities around 6$\cdot$10$^6$cm/s (see Fig. \ref{vel-cza-3D}). It is obvious that
the flow fields in 3D are very different from the ones in 2D, as computed
in KHT, and one expects that this will also effect integral
quantities.

When we stopped this set of computations after 400s, the nuclear
energy generation rate is approximately constant throughout the white
dwarf's atmosphere and a stage of nearly steady hydrogen burning is 
reached. The average nuclear energy generation rate has come down
to a few times 10$^{13}$erg/g/s, missing the conditions for a fast
nova by far. The total energy released by nuclear reactions 
is 5.8$\cdot$10$^{44}$erg which should be compared with the
value of 2.1$\cdot$10$^{45}$erg for our 2D simulation at the 
same time. 

\subsubsection{Temperature evolution, energy generation and convective
mixing}

\begin{figure*}[h]
\begin{center}
\epsfxsize = 8.0cm
\epsfbox{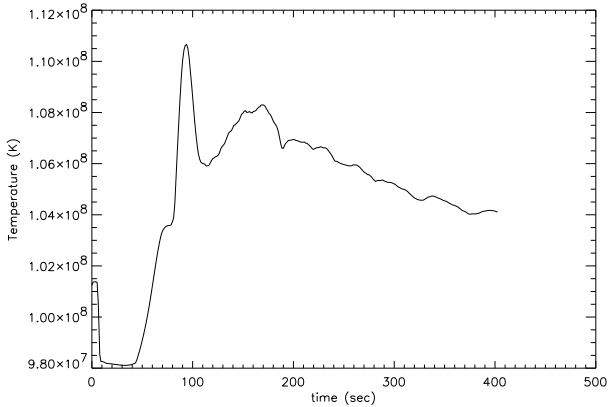}
\caption{\label{temp-3D}
Time evolution of the horizontally averaged temperature in the hottest
layer of the atmosphere for the 3D run with Z = 0.02.}
\end{center}
\end{figure*}

\begin{figure*}[h]
\begin{center}
\epsfxsize = 8.0cm
\epsfbox{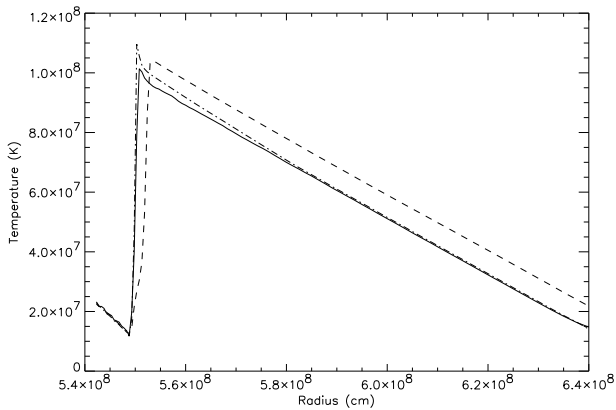}
\caption{\label{avtemp-3D}
Horizontally averaged vertical temperature profiles for the 3D run with
Z = 0.02 at 0 s (solid line),
98 s (dashed-dotted line), and 400 s (dashed line), respectively.}
\end{center}
\end{figure*}

\begin{figure*}[h]
\begin{center}
\epsfxsize = 8.8cm
\epsfbox{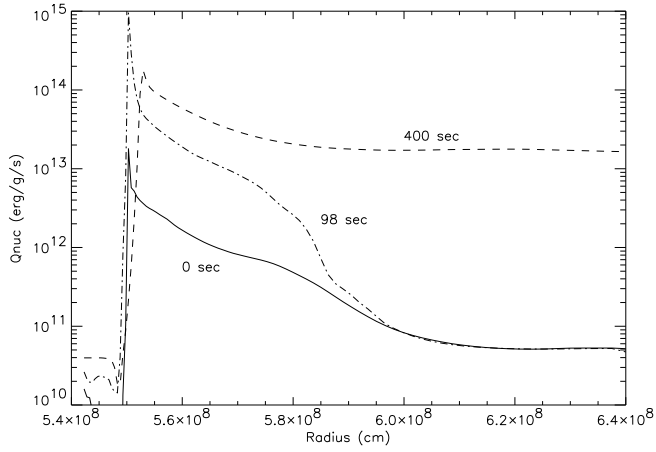}
\caption{\label{qnuc-3D}
Same as Fig. \ref{temp-3D}, but for the horizontally averaged vertical profile of
the energy generation rate.}
\end{center}
\end{figure*}

The temperature evolution of this near-solar metallicity model is
depicted in Figs. \ref{temp-3D} and \ref{avtemp-3D}, and the corresponding 
energy generation
rates are given in Fig. \ref{qnuc-3D}.

In Fig. \ref{temp-3D} we picked the temperature of the hottest zone and plotted it
as a function of time. The first short drop reflects the ignition
phase.
The following rapid rise to the peak value of about
1.1$\cdot$10$^8$K is due to degenerate nuclear burning without
efficient convective energy transport and expansion, as was discussed
in the previous subsection. This phase is followed by some expansion
and adiabatic cooling, accompanied by the onset of more efficient
convective energy transport and mixing. Consequently, the maximum
temperature rises again, but now its peak value is found a bit further
out in the atmosphere. These temperatures are typically 10 to 20 \%
below what we have obtained in our 2D simulations mainly because, as
we shall see, mixing of fresh C and O from the white dwarf into the
hydrogen-rich atmosphere is slower in 3D than in 2D.

Figure \ref{avtemp-3D} shows snapshots of the temperature distribution, averaged over
horizontal planes. With the exception of a clear spike at 98s,
indicating nuclear burning in a well confined surface layer, 
the distributions are very smooth, reflecting the nearly adiabatic
average temperature gradient of a fully convective stratified
atmosphere.

These effects are more clearly seen in
the vertical profiles of the nuclear energy generation rate, again
averaged over horizontal planes, given in Fig. \ref{qnuc-3D}.
For the first about 100s we observe a rapid
increase of the energy generation rate in the layers just on top of
the white dwarf, and convective energy transport is insufficient to
induce fast burning in the outer parts. Towards the end of this
simulations one can see that a stage of nearly steady burning is
reached throughout most of the atmosphere.  

\begin{figure*}[h]
\begin{center}
\epsfxsize = 8.0cm
\epsfbox{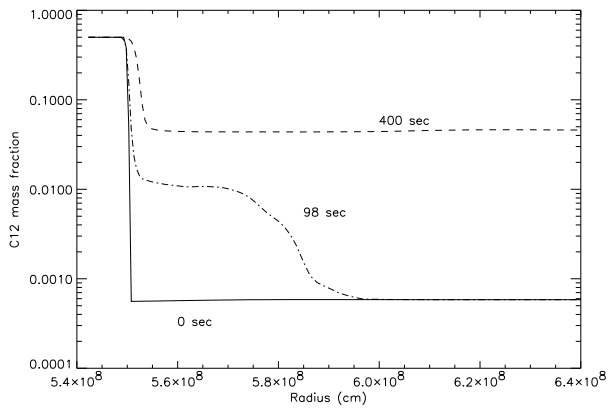}
\caption{\label{avC12-3D}
Same as Fig. 6,  but for the horizontally averaged vertical profiles
of the $^{12}$C mass fraction.}
\end{center}
\end{figure*}

\begin{figure*}[h]
\begin{center}
\epsfxsize = 8.0cm
\epsfbox{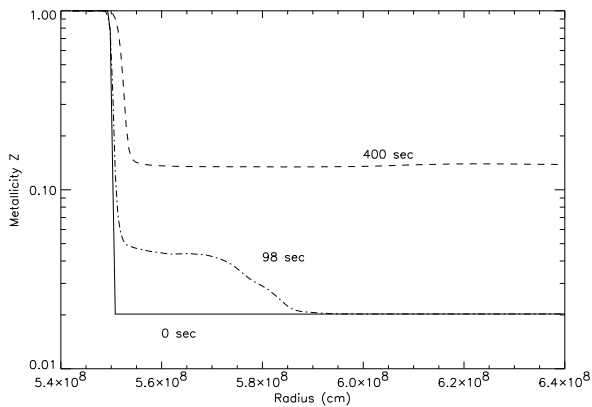}
\caption{\label{avmetall-3D}
Same as Fig. 6, but for the horizontally averaged vertical profiles of
the metallicity.}
\end{center}
\end{figure*}

Finally, in Figs. \ref{avC12-3D} and \ref{avmetall-3D} we show the 
mass fraction of $^{12}$C and
the total metallicity for the same snapshots as in the previous
figures. The sharp discontinuity at 5.5$\cdot$18$^8$cm in the initial
values presents the white dwarf's surface. The initial mass fraction
of $^{12}$C is about 0.06 \% only because steady hydrogen burning 
prior to the TNR has transfered most of the originally solar CNO
isotopes into $^{14}$N, whereas the metallicity is untouched. One can
see in Fig. \ref{avC12-3D} as well as in Fig. \ref{avmetall-3D} that 
fresh $^{12}$C is mixed into
the atmosphere by convective undershoot at a very moderate rate only.
By the time the electron degeneracy is largely removed (after about
100s) mixing has just doubled the metallicity in most of the burning
zones. This is the main reason why we did not obtain a violent
outburst.

\subsubsection{Mass loss and nucleosynthesis}

\begin{figure*}[h]
\begin{center}
\epsfxsize = 8.0cm
\epsfbox{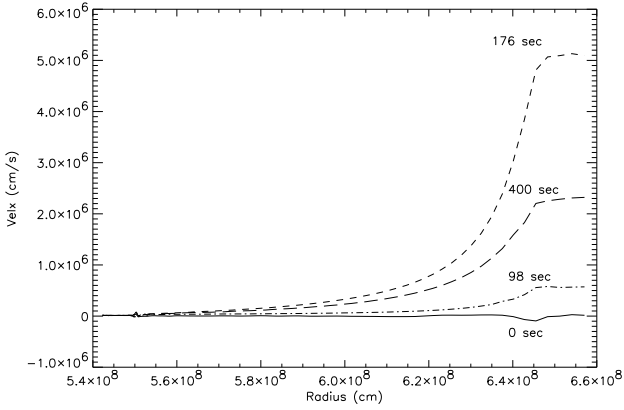}
\caption{\label{avvelx-3D} 
Horizontally averaged vertical profiles of the vertical component of the velocity
at four different times for the 3D simulation with Z = 0.02.}
\end{center}
\end{figure*}

\begin{figure*}[h]
\begin{center}
\epsfxsize = 8.0cm
\epsfbox{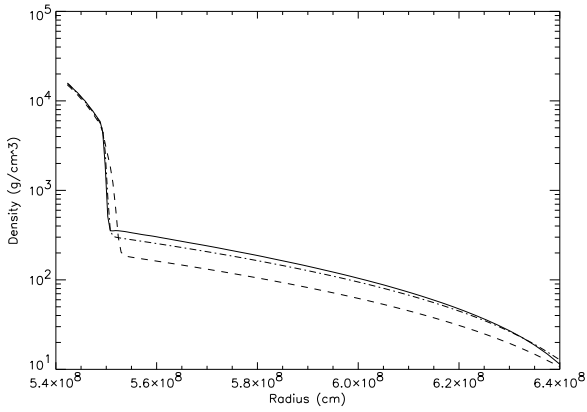}
\caption{\label{avdens-3D}
Horizontally averaged vertical profiles of the density for the run with Z = 0.02
(0 s solid line, 200 s dashed-dotted line, 400 s dashed line, respectively).} 
\end{center}
\end{figure*}

In concluding this section, we want to discuss briefly possible
observable consequences of this model. Peak outflow velocities are
reached after 170s and they amount to 5$\cdot$10$^6$cm/s at a radius of
6400km which is far below the escape velocity from the white dwarf of about 
6$\cdot$10$^8$cm/s (see Fig. \ref{avvelx-3D}). Typically, the
velocities are also a
factor of two below those we obtained in the 2D run at the
same instants of time. Therefore, although we are loosing mass from
our computational grid by outflow (see Fig. \ref{avdens-3D}), there is no direct 
mass ejection from the white dwarf. In contrast, mass loss will happen
in form of a wind from the outer atmosphere which was not included in
our computations.

\begin{figure*}[h]
\begin{center}
\epsfxsize = 7.8cm
\epsfbox{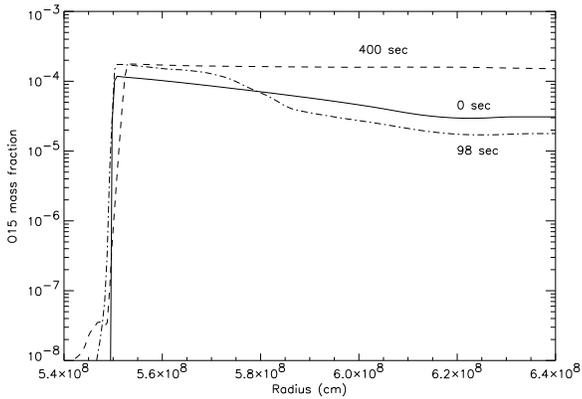}
\caption{\label{avO15-3D} 
Snapshots of the horizontally averaged vertical profiles of the $^{15}$O mass fraction for
the case Z = 0.02.}
\end{center}
\end{figure*}

\begin{figure*}[h]
\begin{center}
\epsfxsize = 7.8cm
\epsfbox{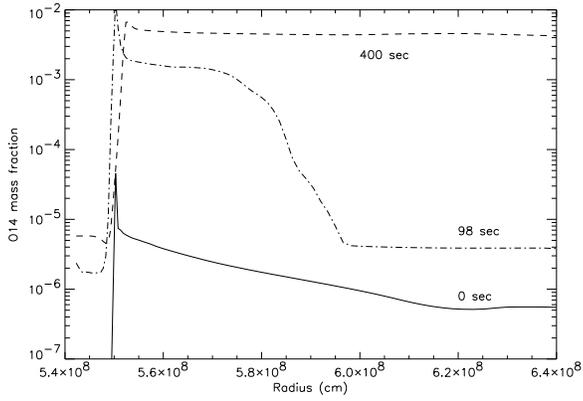}
\caption{\label{avO14-3D}
Same as Fig. \ref{avO15-3D}, but for the $^{14}$O mass fraction.}
\end{center}
\end{figure*}

\begin{figure*}[h]
\begin{center}
\epsfxsize = 7.8cm
\epsfbox{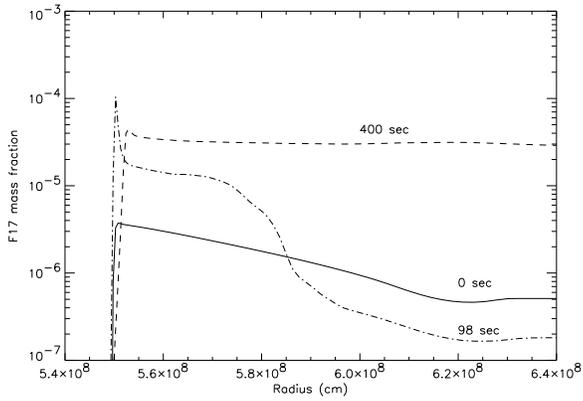}
\caption{\label{avF17-3D}
Same as Fig. \ref{avO15-3D}, but for the $^{17}$F mass fraction.}
\end{center}
\end{figure*}

Conclusions concerning the chemical composition of the wind,
therefore, are rather uncertain. Nevertheless we can obtain rough
estimates by looking at the abundances we find in the computational
domain. As can be seen in Figs. \ref{avO15-3D} through \ref{avF17-3D}, 
after about 400s (in
reality already after 200s) the composition of freshly synthesized
(radioactive) isotopes is homogeneously distributed throughout the
white dwarf's atmosphere. Since we expect that fast convective mixing will
proceed all the way out to optically thin regions one can also expect
that this composition is a fair representation of the matter which is
lost from the white dwarf ejected in the wind. At late times, 
radioactive decay heating by these 
isotopes dominates the energy production in all but the innermost 
layers of the atmosphere, but the overall enrichment of CNO isotopes
is only moderate, probably not more than a factor of five to
six relative to the initial values.

\subsection{A model with Z=0.1}

In one-dimensional nova simulations initial metal abundances largely 
in excess of their solar values are required to drive a fast 
outburst. Since we have demonstrated that self-enrichment of the
atmosphere by convective undershoot and dredge-up of matter from the
white dwarf does not happen during the outburst, we decided to run a
model in which the atmosphere was already enriched from the start.
In order to avoid large inconsistencies (this model was not
evolved in the same way as the low-metallicity one), we increased Z to
0.1 only and constructed a new equilibrium model which had the same
temperature structure as before. Burning was then ignited by raising
the temperature by 1\% in 20 zones of the bottom layer of the accreted
atmosphere, and all other properties of the model were identical to
the previous one. Of course, an enrichment by a factor of five is
still below what is commonly assumed in 1D simulations but one might
hope for more violent burning and, therefore, also faster mixing of C
and O already in this case, if convective motions are treated properly.
However, as we shall demonstrate, this is not the case. We find that
the evolution proceeds considerably faster than in the low-metallicity
case, but the final outcome is not too different in both cases. 

\begin{figure*}[h]
\begin{center}
\epsfxsize = 8.0cm
\epsfbox{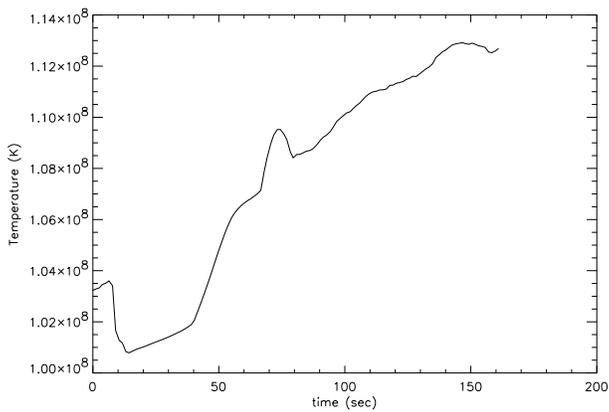}
\caption{\label{temp-3D_2} 
Temperature in the hottest shell of the atmosphere for the 3D run with
an initial metallicity of Z = 0.1.}
\end{center}
\end{figure*}

\begin{figure*}[h]
\begin{center}
\epsfxsize = 8.0cm
\epsfbox{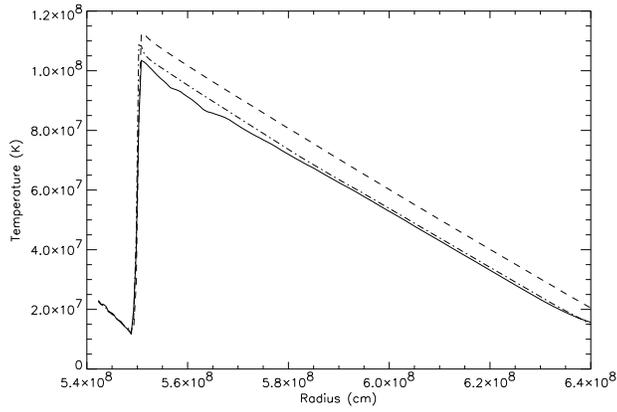}
\caption{\label{avtemp-3D_2}
Snapshots of the horizontally averaged vertical profiles of the
temperature for the run with Z = 0.1 initially,  at 0 s (solid line),
78 s (dashed-dotted line) and 160 s (dashed line).}
\end{center}
\end{figure*}

\begin{figure*}[h]
\begin{center}
\epsfxsize = 8.0cm
\epsfbox{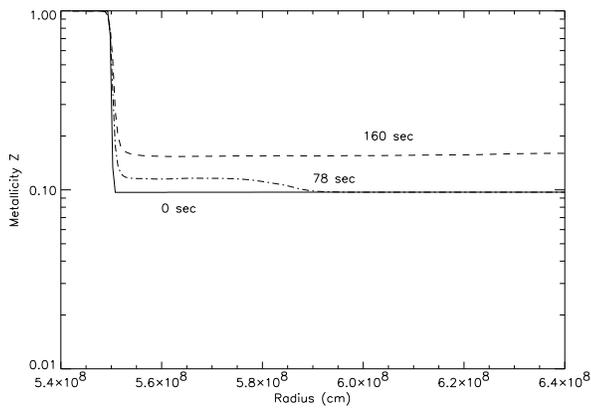}
\caption{\label{avmetall-3D_2}
Same as Fig. \ref{avtemp-3D_2}, but for  the metallicity.}
\end{center}
\end{figure*}

Since the flow patterns look very similar to those shown in Fig. \ref{flow-3D},
with the exception that already after about 80s velocities above
10$^7$cm/s are observed, and after about 150s peak velocities reach 
2$\cdot$10$^7$cm/s, nearly a factor of 2 higher than in the
low-metallicity run. The second effect is that peak temperatures are
now a bit higher (see Figs. \ref{temp-3D_2} and \ref{avtemp-3D_2}, to 
be compared with Figs. \ref{temp-3D} and \ref{avtemp-3D}),
and also the energy generation rate exceeds the previous one, 
but not by much. After 160s a state of burning is reached which is nearly
indistinguishable from the low-metallicity case at 400 seconds. Even the metallicity in
the atmosphere (Fig. \ref{avmetall-3D_2}) is only moderately higher than before.

We therefore conclude that, although originally higher CNO-abundances
speed up the thermonuclear burning, going to 5 times solar metallicity
is still insufficient for a fast nova. The reason seems to be that,
despite the fact that the consumption of hydrogen is faster in the
beginning, the time scale for the temperature rise is still too long,
i.e. on the order of 100s. This, in turn, means that, as before, the 
atmosphere begins to expand and cool once the temperature exceeds 
10$^8$K, much too low for a fast nova. Very fast mixing 
of white dwarf material into the atmosphere during the early phase of
the TNR could, in principle, overcome this problem but we do not find
such a mixing. In contrast, the 3D simulations mix even less than the 2D
models of KHT.

\section{Summary and conclusions}

We have presented the first 3-dimensional simulations of thermonuclear
models of classical novae. We have shown that the results differ, in
several aspects, considerably from those of previous 
2-dimensional simulations and we conclude that
this and similar problems have to be carried out in 3D in order to give
reliable results. The main reason is that the flow fields in 3D are
very different from those in 2D and, therefore, all quantities which
depend on them, such as convective energy transport and mixing, will
also differ in both cases. For example, in 3D we find, not
unexpectedly, more power in small scale motions than in our previous
2D simulations which leads to less convective undershoot into the
white dwarf's surface and consequently less dredge-up of C and O.
This, in turn, makes explosive hydrogen burning less violent,
weakening the chances for getting a fast nova that way.

Therefore we arrive at the conclusion that self-enrich\-ment 
of the accreted atmosphere with C and O during the
outburst is very unlikely, if not impossible, and our simulations rule
out one of the suggestions for changing a slow nova into a fast one.
A more likely solution to this problem seems to be large enrichment
prior to the outburst by either shear-induced instabilities at the
interface between the white dwarf and its atmosphere (like in an
accretion belt) or some kind of diffusive or convective 
mixing during the long quiet accretion phase.

Finally, we want to point out that the type of numerical simulations
we have performed leave room for improvements. Firstly, as was mentioned
already in the introduction, in our numerical method mixing on small
scales is due to numerical diffusion. It is difficult to estimate
quantitatively its effect. It appears to be safer to include mixing by small
scale turbulence explicitly by a subgrid model, as was done
for the propagation of nuclear flames by Niemeyer \& Hillebrandt
(1995). Secondly, the kind of problems we have tackled here with brute
force, namely a problem in which the typical velocities are far below
the sound velocity, should be better approached by means of (nearly
incompressible) implicit hydrodynamics  schemes which are presently being
developed. It is obvious that one would like to explore a larger
fraction of the parameter space, but this can only be done with more
efficient codes.

{\it Acknowledgments.} The authors are grateful for many enlightening
discussions with Stanford E. Woosley, Jens C. Niemeyer and Hans Ritter. 
They thank Ami Glasner and Eli Livne for supplying the initial model,
Ewald M\"uller for an earlier version of the PROMETHEUS code, and Rudi
Fischer and Jakob Pichlmeier for their help in preparing a parallel version
of the code. They also thank the staff members of the Rechenzentrum Garching
for their support during the computations. This work was supported in part 
by NASA under Grant NAG 5-3076, by DOE under contract No. B341495 at
the University of Chicago, and by the Deutsche Forschungsgemeinschaft
under Grant Hi 534/3-1.
The computations were performed at the Rechenzentrum Garching on a
Cray T3E.
%         
%

%%%%%%%%%%%%%%%%%%%%%%%%%
%%%%%%%  FIGURES  %%%%%%%
%%%%%%%%%%%%%%%%%%%%%%%%%

% FIGURE 1

\begin{figure*}[t]
\begin{center}
%\begin{tabular}{cc}
%\epsfxsize = 6.8cm
%\epsfbox{../../results/3D/vel_slices/Vel_NOVAcabm.ps}
%&
%\epsfxsize = 6.8cm
%\epsfbox{../../results/3D/vel_slices/Vel_NOVAcada.ps} \\
%a) & b) \\ \\
%\epsfxsize = 6.8cm
%\epsfbox{../../results/3D/vel_slices/Vel_NOVAcbem.ps}
%&
%\epsfxsize = 6.8cm
%\epsfbox{../../results/3D/vel_slices/Vel_NOVAcbmm.ps} \\
%c) & d) \\ \\
%\epsfxsize = 6.8cm
%\epsfbox{../../results/3D/vel_slices/Vel_NOVAccza.ps}
%&
%\epsfxsize = 6.8cm
%\epsfbox{../../results/3D/vel_slices/Vel_NOVAcelm.ps} \\
%e) & f)
%\end{tabular}
\caption{\label{flow-3D} Snapshots of the
velocity field of the three-dimensional run with an initial metallicity
of Z = 0.02. Shown are vertical slices through the calculation domain.
The absolute values of the velocity at each point are color coded.
Note that the ``radial'' coordinate is logarithmic above
6500 km.}
\end{center}
\end{figure*}

% FIGURE 2

\begin{figure*}[t]
\begin{center}
%\begin{tabular}{cc}
%\epsfxsize = 7.4cm
%\epsfbox{../../results/3D/RENDER/vel/vel_ata/vel_1.5E6_ata.ps}
%&
%\epsfxsize = 7.4cm
%\epsfbox{../../results/3D/RENDER/vel/vel_ata/vel_1.0E5_ata.ps} \\
%Iso-surface $v = 1.5 \cdot 10^6$ cm/s & Iso-surface $v = 1.0 \cdot 10^5$ cm/s
%\end{tabular}
\caption{\label{vel-ata-3D}
Iso-surfaces of the absolute value of the
velocity for the 3D simulation with an initial
metallicity of Z = 0.02 at {\bf $t$ = 50 seconds}.
The (800km)$^3$ cube shown in this figure is only a fraction
of the total computational domain. Two distinctly different
convective layers can be seen: one of high velocity near to the
surface of the white dwarf, and a second one of low velocities
in the outer atmosphere.}
\end{center}
\end{figure*}

% FIGURE 3

\begin{figure*}[t]
\begin{center}
%\begin{tabular}{cc}
%\epsfxsize = 7.4cm
%\epsfbox{../../results/3D/RENDER/vel/vel_bmm/vel_1.0E7_bmm.ps}
%&
%\epsfxsize = 7.4cm
%\epsfbox{../../results/3D/RENDER/vel/vel_bmm/vel_4.0E6_bmm.ps} \\
%Iso-surface $v = 1.0 \cdot 10^7$ cm/s & Iso-surface $v = 4.0 \cdot 10^6$ cm/s \\ \\
%\epsfxsize = 7.4cm
%\epsfbox{../../results/3D/RENDER/vel/vel_bmm/vel_8.0E5_bmm.ps}
%&
%\epsfxsize = 7.4cm
%\epsfbox{../../results/3D/RENDER/vel/vel_bmm/vel_2.0E5_bmm.ps} \\
%Iso-surface $v = 8.0 \cdot 10^5$ cm/s & Iso-surface $v = 2.0 \cdot 10^5$ cm/s \\ \\
%\end{tabular}
\caption{\label{vel-bmm-3D} Same as Fig.2, but at {\bf $t$ = 100 seconds}.
It can be seen that typical convective velocities still decrease
with increasing distance from the white dwarf's surface, but the
different layers begin to mix. Those close to the surface are heavily
stirred by nuclear energy generation.}
\end{center}
\end{figure*}

% FIGURE 4

\begin{figure*}[t]
\begin{center}
%\begin{tabular}{cc}
%\epsfxsize = 7.4cm
%\epsfbox{../../results/3D/RENDER/vel/vel_cza/vel_1.0E7_cza.ps}
%&
%\epsfxsize = 7.4cm
%\epsfbox{../../results/3D/RENDER/vel/vel_cza/vel_8.0E6_cza.ps} \\
%Iso-surface $v = 1.0 \cdot 10^7$ cm/s & Iso-surface $v = 8.0 \cdot 10^6$ cm/s \\ \\
%\epsfxsize = 7.4cm
%\epsfbox{../../results/3D/RENDER/vel/vel_cza/vel_4.0E6_cza.ps}
%&
%\epsfxsize = 7.4cm
%\epsfbox{../../results/3D/RENDER/vel/vel_cza/vel_1.0E6_cza.ps} \\
%Iso-surface $v = 4.0 \cdot 10^6$ cm/s & Iso-surface $v = 1.0 \cdot 10^6$ cm/s \\ \\
%\end{tabular}
\caption{\label{vel-cza-3D}
Same as Fig.2, but at {\bf $t$ = 200 seconds}.
The convective eddies no longer show a clear structure. The entire
computational domain is well stirred and covered by eddies of all
sizes and different velocities.}
\end{center}
\end{figure*}

% FIGURES 5, 6 and 7


% FIGURES 8, 9, 10 and 11



% FIGURES 12, 13 and 14


% FIGURES 15, 16 and 17


\end{document}